\documentclass[11pt]{article}
\usepackage{latexsym}
\usepackage{graphicx}
\usepackage{amsfonts,amssymb}

\begin{document}

\title{A new mathematical representation of Game Theory I}
\author{Jinshan Wu
\\ Department of Physics, Simon Fraser University, Burnaby, B.C.
Canada, V5A 1S6}

\maketitle

\begin{abstract}
In this paper, we introduce a framework of new mathematical
representation of Game Theory, including static classical game and
static quantum game. The idea is to find a set of base vectors in
every single-player strategy space and to define their inner
product so as to form them as a Hilbert space, and then form a
Hilbert space of system state. Basic ideas, concepts and formulas
in Game Theory have been reexpressed in such a space of system
state. This space provides more possible strategies than
traditional classical game and traditional quantum game. So
besides those two games, more games have been defined in different
strategy spaces. All the games have been unified in the new
representation and their relation has been discussed. General Nash
Equilibrium for all the games has been proposed but without a
general proof of the existence. Besides the theoretical
description, ideas and technics from Statistical Physics, such as
Kinetics Equation and Thermal Equilibrium can be easily
incorporated into Game Theory through such a representation. This
incorporation gives an endogenous method for refinement of
Equilibrium State and some hits to simplify the calculation of
Equilibrium State. The more privileges of this new representation
depends on further application on more theoretical and real games.
Here, almost all ideas and conclusions are shown by examples and
argument, while, we wish, lately, we can give mathematical proof
for most results.
\end{abstract}


Pacs: 02.50.Le, 03.67.-a, 03.65.Yz, 05.20.-y, 05.30.-d

\newpage

\section{Introduction}
Game Theory\cite{Gamecourse, Gamecourse2} is a subject used to
predict the strategy of all players in a game. The simplest game
is static and non-cooperative game, which describe by payoff
function $G$, a linear mapping from strategy space $S^{1}\times
S^{2}\times\cdots S^{N}$ to N-dimension real space
$\mathbb{R}^{N}$. For a game in mixture strategy space, in which
every player uses mixture strategy $\vec{P}^{i}$, a probability
distribution on strategy set $S^{i}$, Nash Theorem proves that
there is always some mixture-strategy equilibrium points, on which
no player has the willing to make an independent change. Therefor,
such equilibrium points can be regarded as a converged points (or
at least fixed points) of the system, then as the end state of all
players.

On the other hand, Quantum Game Theory\cite{Qgamereview1,
Qgamereview2} has been proposed as a quantum version of Game
Theory\cite{meyer,jens}. A typical two-player quantum game is
defined as $\Gamma = \left(\mathcal{H}, \rho_{0}, S_{A}, S_{B},
P_{A}, P_{B}\right)$, in which $\mathcal{H}$ is the Hilbert space
of state of one quantum object, like a photon or electron. Such a
quantum object plays an important role in Quantum Game Theory. The
quantum strategy set $S_{A}$ or $S_{B}$ usually is defined as a
set of unitary operator on state space $\mathcal{H}$, or sometimes
subspace of $\mathcal{H}$. It's believed that because quantum
strategy space is usually larger than the corresponding classical
strategy space, one can make use of such advantage of quantum
strategy to make money over classical player. Generally speaking,
a quantum game is not only a quantum version of a classical game.
It describes the situation that every player plays game in a
strategy space of quantum operators, which transform the state of
a quantum object. Probably, someday, some original quantum games
can be found alone this direction.

However, both of Classical Game Theory and Quantum Game Theory are
expressed in single player strategy space, so the payoff function
is a $\left(0, N\right)$ tensor $G^{i}\left(s^{1}, s^{2}, \dots,
s^{N}\right)$, mapping a combination of $N$ single-player
strategies onto a real number. On the contrary, in Quantum
Statistical Mechanics, a matrix form of Hamiltonian $H$ is used in
a any-particle case, and the form of density matrix of equilibrium
state is always $\rho = e^{-\beta H}$. So a system-level
description will unify our formulas for $N$-player game, and then
maybe improve our understanding and calculation.

Starting from such an idea, in this paper, we construct a
systematical way to reexpress everything into system-level
description, including system state and its space, payoff matrix
on system space, and reduced single-player payoff matrix. Then
Canonical Quantum Ensemble distribution is used to describe system
equilibrium state. So ideas and technics from Statistical Physics
can be easily applied onto Game Theory. Such application implies a
probability that a Kinetics Equation can be used to describe an
evolution that a system ends at the equilibrium state starting
from an arbitrary distribution. Because the traditional Game
Theory only cares about the macro-equilibrium state, the Kinetics
Equation approach is just pseudo-dynamical equation leading to the
equilibrium state. The dynamical process itself might be
meaningless.

Furthermore, besides providing a new pseudo-dynamical approach,
the distribution function description has its own meaning. In Game
Theory, maybe general for all economical subjects, usually it's
supposed that even the difference between two choices is very
small, the high-value one is chosen. This is unnatural when the
difference is smaller than the resolution of human decision.
Therefor, we use a distribution function to replace the
maximum-point solution. This means player $i$ will choose strategy
$s^{i}_{\mu}$ with probability $e^{\beta E^{i}(s^{i}_{\mu})}$,
even there are another strategy can make more money. Here, $\beta$
is the meaning of average resolution level, or in Statistical
Physics, the average noise level. Unfortunately, although the
ensemble description is the second topic of this paper, only some
special case study has been investigated. A general form for any
game is still waiting for more exploration.

Section $\S$\ref{framework} constructs the new representation for
classical and quantum game. Section $\S$\ref{statphys} use
ensemble distribution and pseudo-dynamical approach to study the
equilibrium state in this new representation. Discussion of
relation between our new representation and quantum, and classical
game is included in section $\S$\ref{framework}. A lot of
questions are pointed out in the discussion section
($\S$\ref{discussion}). Section $\S$\ref{conclusion} is a short
summary of the conclusions we have reached.

\section{Mathematical Structure: Strategy Space, State Density Matrix and Payoff Matrix}
\label{framework}

Strategy set can be continuous and discrete, and this will effect
the mathematical form of all variables, such as the state of
player $i$ is $p(s^{i})$ or
$p^{i}_{\mu}\delta(s^{i}-s^{i}_{\mu})$, and $G$ will be
integrations or matrixes. In order to compare with the
Mathematical form of Quantum Mechanics and point out the
similarity, and to unify Classical Game Theory and Quantum Game
Theory, here we use the discrete strategy, although the
corresponding transformation of all ideas and formulas is quite
straightforward. Most of our formulas and results can be
generalized into $N$-player and $\left(\prod_{i=1}^{N}
L_{i}\right)$-strategy game, so for simplicity of expressions, at
most time, a $2$-player and $\left(L_{1}\times
L_{2}\right)$-strategy game is used as our object.

\subsection{The new representation of static classical game}
For a $N$-player game, we suppose the strategy space of player $i$
is $S^{i}=\left\{s^{i}_{1}, s^{i}_{2}, \cdots,
s^{i}_{L_{i}}\right\}$. The state of player $i$ is
$\left|P^{i}\right>\equiv\left(p^{i}_{1}, p^{i}_{2}, \cdots,
p^{i}_{L_{i}}\right)^{T}$ and $\sum_{\mu=1}^{L_{i}}p^{i}_{\mu} =
1$. The payoff function of player $i$ is a $\left(0,
N\right)$-tensor
--- a $N$-linear operator,
\begin{equation}
E^{i}(P^{1}, P^{2}, \dots, P^{N} ) =
G^{i}\left(\left|P^{1}\right>, \left|P^{2}\right>, \cdots,
\left|P^{N}\right> \right).
\label{singlepayoff}
\end{equation}
Specially, for a 2-player game, $G^{i}$ can be written as a matrix
($\left(0,2\right)$-tensor) so that
\begin{equation}
E^{i}(P^{1},P^{2})=\left<P^{1}\right|G^{i}\left|P^{2}\right>,
\label{singlepayoff2}
\end{equation}
in which $G^{i}$ is $L_{1}\times L_{2}$ matrix, not necessary a
square one. So a classical game is
\begin{equation}
\Gamma_{C} = \left(\left\{S^{i}\right\},
\left\{G^{i}\right\}\right),
\label{oldclassicalgame}
\end{equation}
A general vector in $S^{i}$ can be defined as
\begin{equation}
\left|P^{i}\right> = \sum_{\mu =
1}^{L_{i}}p^{i}_{\mu}\left|s^{i}_{\mu}\right>,
\label{oldsinglestatevector}
\end{equation}
in which $\left\{s^{i}_{\mu}\right\}$ is the base vector set of
strategy space $S^{i}$. Or in traditional language of Game Theory,
it's a set of all the pure strategies of player $i$.

Inspired by the application of Hilbert Space in Quantum Mechanics,
now we introduce two ideas into Game Theory. First, to redefine
the strategy space of single player as a Hilbert Space. Second, to
use a system state to replace the single player state. Then, at
the same time, a new form of payoff function is required to be
equivalently defined on the system state.

A single-player state vector $\left|P^{i}\right>$ is written in a
new form as
\begin{equation}
\rho^{i} = \sum_{\mu = 1}^{L_{i}}
p^{i}_{\mu}\left|s^{i}_{\mu}\right>\left<s^{i}_{\mu}\right|.
\label{newsinglestatevector}
\end{equation}
It's density matrix form of a mixture state, because a classical
strategy of player $i$ is to use strategy $s^{i}_{\mu}$ with
probability $p^{i}_{\mu}$. The difference between
equ(\ref{oldsinglestatevector}) and
equ(\ref{newsinglestatevector}) can be regarded as just to replace
$\left|s^{i}_{\mu}\right>$ with
$\left|s^{i}_{\mu}\right>\left<s^{i}_{\mu}\right|$. The reason of
such replacement will be clear when we do it on quantum game.
Actually, using density matrix to describe mixture state is a
approach in Quantum Mechanics. A system state of all players is
defined as
\begin{equation}
\rho^{s} = \prod_{i=1}^{N}\rho^{i}.
\label{newsystemstatevector}
\end{equation}
A typical form of system state of a $2$-player (player $1$ and
player $2$) and $2$-strategy (strategy $\left(\alpha,
\beta\right)$ and strategy $\left(\mu, \nu\right)$) classical game
looks like
\begin{equation}
\rho^{s} =
p^{1}_{\alpha}p^{2}_{\mu}\left|\alpha\mu\right>\left<\alpha\mu\right|
+
p^{1}_{\alpha}p^{2}_{\nu}\left|\alpha\nu\right>\left<\alpha\nu\right|
+
p^{1}_{\beta}p^{2}_{\mu}\left|\beta\mu\right>\left<\beta\mu\right|
+
p^{1}_{\beta}p^{2}_{\nu}\left|\beta\nu\right>\left<\beta\nu\right|.
\label{specificcase}
\end{equation}
In fact, the base vector set of Hilbert space of system state of
$N$ players can be defined as direct product of single player base
vector as
\begin{equation}
\left|S_{\vec{\mu}}\right> := \left|s^{1}_{\alpha}, s^{2}_{\beta},
\dots, s^{N}_{\gamma} \right> = \prod_{i =
1}^{N}\left|s^{i}_{\mu}\right>.
\label{systembase}
\end{equation}
Then from equ(\ref{newsinglestatevector}) and
equ(\ref{newsystemstatevector}), it can be proved that a system
state have the form as
\begin{equation}
\rho^{s} =
\sum_{\vec{\mu}}P_{S_{\vec{\mu}}}\left|S_{\vec{\mu}}\right>\left<S_{\vec{\mu}}\right|,
\label{systemdensity}
\end{equation}
in which
\begin{equation}
P_{S_{\vec{\mu}}} = \prod_{i=1}^{N}p^{i}_{\mu}.
\end{equation}
One can compare this general form with the specific one of
$2\times2$ game, equ(\ref{specificcase}). Sometimes, we neglect
the subindex and denote $\left|S_{\vec{\mu}}\right>$ as
$\left|S\right>$. In such situation, we should notice that a
capital $S$ denote a general system base vector.

Such replacement provides a probability to use pure strategy other
than the traditional classical pure and mixture strategy. We will
discuss this in section $\S$\ref{quantizedclassical}. Now we try
to transform payoff function $G^{i}$ into system-level form while
the invariant condition is equ(\ref{singlepayoff}). In a density
matrix form, the formula used to calculate the payoff is
\begin{equation}
E^{i}(\rho^{s})=Tr(\rho^{s}H^{i}) = \sum_{S}
\left<S\right|\rho^{s}H^{i}\left|S\right>,
\label{systempayoff}
\end{equation}
The solution of equ(\ref{singlepayoff}) and
equ(\ref{systempayoff}) gives the relation between $H^{i}$ and
$G^{i}$. Since those two equations should give the same value for
any state, we can choose the system state as a pure strategy, or
in our language, the base vector of system state. Let's denote
$\left|S_{0}\right> = \left|s^{1}_{0}, s^{2}_{0},\dots,
s^{N}_{0}\right>$, which means every player choose a pure strategy
$s^{i}_{0}$, then $P^{i}_{0} = \delta_{s^{i}s^{i}_{0}}$. Then
equ(\ref{systempayoff}) give us
\[
E^{i}(\left|S_{0}\right>\left<S_{0}\right|) = \sum_{S}
\left<S|S_{0}\right>\left<S_{0}\right|H^{i}\left|S\right>  =
\left<S_{0}\right|H^{i}\left|S_{0}\right>
\]
So
\begin{equation}
\left<S_{0}\right|H^{i}\left|S_{0}\right> =
G^{i}\left(\left|s^{1}_{0}\right>, \left|s^{2}_{0}\right>, \cdots,
\left|s^{N}_{0}\right> \right).
\label{relation}
\end{equation}
The diagonal elements of $H^{i}$ can be calculated explicitly. And
for our general system density matrix as equ(\ref{systemdensity}),
only the diagonal terms effect the payoff value $E^{i}$, all
others can defined as zero. For example, $H^{i}$ of a 2-player
game is
\begin{equation}
H^{i} =\sum_{\alpha\beta,\mu\nu}
G^{i}_{\alpha\mu}\delta_{\alpha\beta}\delta_{\mu\nu}
\left|\alpha,\mu\right>\left<\beta,\nu\right|.
\label{relation2}
\end{equation}
This means $H^{i}$ is diagonal matrix.

\subsection{Prisoner's Dilemma as an example}
Before we continue our further discussion, let's use one example
to present our abstract Mathematics and to compare the traditional
and new from of state vector and payoff function. The traditional
payoff function of Prisoner's Dilemma is
\[
\begin{array}{|c|c|c|}
\hline & Cooperate & Defect \\\hline Cooperate & -2,-2 & -5,0
\\\hline
Defect & 0,-5 & -4,-4 \\\hline
\end{array}
\]
Then
\[
G^{1}=\left[\begin{array}{cc} -2 & -5
\\0 & -4
\end{array}\right], G^{2}=\left[\begin{array}{cc}
-2 & 0
\\-5 & -4
\end{array}\right].
\]
The traditional state vectors are
\[
\begin{array}{ll}
\left|P^{1}\right>^{old} = p^{1}_{c}\left|C\right> +
p^{1}_{d}\left|D\right>, & \left|P^{2}\right>^{old} =
p^{2}_{c}\left|C\right> + p^{2}_{d}\left|D\right>
\end{array}.
\]
By substituting the above two equations into
equ(\ref{singlepayoff2}), we get
\begin{equation}
\begin{array}{lll} E^{1} & = & \left[\begin{array}{cc}p^{1}_{c} &
p^{1}_{d}\end{array}\right]\left[\begin{array}{cc} -2 & -5
\\0 & -4
\end{array}\right]\left[\begin{array}{c}p^{2}_{c}\\p^{2}_{d}
\end{array}\right]
\\
\\ & = & -2p^{1}_{c}p^{2}_{c}-5p^{1}_{c}p^{2}_{d}+0\cdot
p^{1}_{d}p^{2}_{c}-4p^{1}_{d}p^{2}_{d}
\end{array}
\label{oldearn}
\end{equation}
By the new notations, state of player $1$ is
\[
\rho^{1} = p^{1}_{c}\left|C\right> \left<C\right| +
p^{1}_{d}\left|D\right> \left<D\right|.
\]
and system state is
\[
\rho^{s} = p^{1}_{c}p^{2}_{c}\left|CC\right> \left<CC\right| +
p^{1}_{c}p^{2}_{d}\left|CD\right> \left<CD\right| +
p^{1}_{d}p^{2}_{c}\left|DC\right> \left<DC\right| +
p^{1}_{d}p^{2}_{d}\left|DD\right> \left<DD\right|,
\]
or in matrix form,
\[
\rho^{s} = \left[\begin{array}{cccc}p^{1}_{c}p^{2}_{c} & 0 & 0 & 0
\\0 & p^{1}_{c}p^{2}_{d} & 0 & 0
\\0 & 0 & p^{1}_{d}p^{2}_{c} & 0
\\0 & 0 & 0 & p^{1}_{d}p^{2}_{d}
\end{array}\right].
\]
Then from equ(\ref{relation2}), we know the new payoff function,
\[
H^{1}=\left[\begin{array}{cccc}
-2 & 0 & 0 & 0
\\0 & -5 & 0 & 0
\\0 & 0 & 0 & 0
\\0 & 0 & 0 & -4
\end{array}\right], H^{2}=\left[\begin{array}{cccc}
-2 & 0 & 0 & 0
\\0 & 0 & 0 & 0
\\0 & 0 & -5 & 0
\\0 & 0 & 0 & -4
\end{array}\right].
\]
We can check it by substituting into equ(\ref{systempayoff}) as,
\begin{equation}
E^{1} = Tr\left(\rho^{s}H^{1}\right) =
-2p^{1}_{c}p^{2}_{c}-5p^{1}_{c}p^{2}_{d}+0\cdot
p^{1}_{d}p^{2}_{c}-4p^{1}_{d}p^{2}_{d},
\label{newearn}
\end{equation}
which is the same value with equ(\ref{oldearn}). So the new
representation includes all the information in the traditional
notation, however, more complex it seems. But such complexity
brings some other benefit including Equilibrium State calculation
and generalization into Quantum Game Theory.

\subsection{Quantized classical game: expanded strategy space}
\label{quantizedclassical}

Till now, since the classical strategy is a mixture strategy of
the base vector (strategy), we always use density matrix to
represent a single player state or a system state, such as in
equ(\ref{newsinglestatevector}) and
equ(\ref{newsystemstatevector}). Now we ask the question that
what's the pure state of strategy (but other than the classical
pure strategy) means in Game Theory? A pure strategy vector of
player $i$ in our representation is
\begin{equation}
\left|P^{i}\right>^{pure} =
\sum_{\mu}^{L_{i}}x^{i}_{\mu}\left|\mu\right>,
\end{equation}
Therefor, the density matrix of such a pure state is
\begin{equation}
\rho^{i,pure} = \left|P^{i}\right>^{pure}\left<P^{i}\right|^{pure}
=\sum_{\mu\nu}^{L_{i},L_{i}}x^{i}_{\mu}\bar{x}^{i}_{\nu}\left|\mu\right>\left<\nu\right|.
\label{puredensity}
\end{equation}
The density matrix of a pure state has off-diagonal elements while
the classical mixture density matrix has only the diagonal
elements. It seems that pure strategies expand the strategy space.
Whether it has significant result in Game Theory or not? Comparing
equ(\ref{puredensity}) with equ(\ref{newsinglestatevector}), if we
suppose
\begin{equation}
\left|x^{i}_{\mu}\right|^{2} = x^{i}_{\mu}\bar{x}^{i}_{\mu} =
p^{i}_{\mu},
\end{equation}
that every diagonal element of pure density equals to
corresponding one of mixture density matrix, then those two states
will have similar meaning. Let's use the Prisoner's Dilemma as an
example again to check if they will give different payoffs.
Although we still can follow the calculation of mixture state by
density matrix method as in equ(\ref{systempayoff}), there is an
equivalent but much simpler formula for pure state calculation,
\begin{equation}
E^{i,pure} = \left<S\right|H^{i}\left|S\right>.
\end{equation}
Where $\left|S\right>$ is a pure state vector defined direct
product of single player state as
\begin{equation}
\left|S\right> =
\left|P^{1}\right>\left|P^{2}\right>\dots\left|P^{N}\right>
:=\left|P^{1},P^{2},\dots,P^{N}\right>.
\end{equation}
Specially, for Prisoner's Dilemma, the system state vector is
\[
\left|S\right> = x^{1}_{c}x^{2}_{c}\left|CC\right> +
x^{1}_{c}x^{2}_{d}\left|CD\right> +
x^{1}_{d}x^{2}_{c}\left|DC\right> +
x^{1}_{d}x^{2}_{d}\left|DD\right> .
\]
Combined with new payoff matrix $H^{1}$,
\[
\begin{array}{lll}
E^{i,pure} & = & \left[
\begin{array}{cccc}\bar{x}^{1}_{c}\bar{x}^{2}_{c} & \bar{x}^{1}_{c}\bar{x}^{2}_{d} & \bar{x}^{1}_{d}\bar{x}^{2}_{c} &
\bar{x}^{1}_{d}\bar{x}^{2}_{d}\end{array}\right]
\left[\begin{array}{cccc} -2 & 0 & 0 & 0
\\0 & -5 & 0 & 0
\\0 & 0 & 0 & 0
\\0 & 0 & 0 & -4
\end{array}\right]\left[
\begin{array}{c}x^{1}_{c}x^{2}_{c}
\\ x^{1}_{c}x^{2}_{d}
\\ x^{1}_{d}x^{2}_{c}
\\ x^{1}_{d}x^{2}_{d}
\end{array}\right]
\\
& = & -2p^{1}_{c}p^{2}_{c}-5p^{1}_{c}p^{2}_{d}+0\cdot
p^{1}_{d}p^{2}_{c}-4p^{1}_{d}p^{2}_{d}
\end{array}
\]
the same value as equ(\ref{oldearn}) and equ(\ref{newearn}).

Pure state $\rho^{s,pure}$ equals to the diagonal term plus some
off-diagonal elements. The same payoff value implies that in our
situation, only the diagonal term makes sense. In fact, generally,
it's because of the diagonal property of $H^{i}$ . From
equ(\ref{systempayoff}) and equ(\ref{relation}), we know
\begin{equation}
E^{i} = \sum_{S} \left<S\right| \rho H^{i} \left|S\right> =
\sum_{S} \left<S\right|\rho\left|S\right> H_{ss}^{i}= \sum_{S}
\rho_{ss}H_{ss}^{i}.
\end{equation}
It means that even $\rho^{s,pure}$ has off-diagonal elements, only
the diagonal parts effect the payoff. In one word, the
mathematical form of vectors in Hilbert space, or the equivalent
density matrix form, brings nothing new into Game Theory but an
equivalent mathematical representation. System state can be a pure
state or a mixture state, but since the payoff matrix is diagonal,
they make no difference. The quantization of Classical Game is a
new game only when {\emph{both density matrix and payoff matrix
have off-diagonal elements}}. Although density matrix $\rho^{s}$
can have the off-diagonal terms, the relation of
equ(\ref{relation}) between the new payoff matrix and the
traditional one guarantees that $H^{i}$ can only have the diagonal
term. So this quantization condition has no classical meaning in
Game Theory.

Now the question is if we quantize it anyway, what's the meaning?
Is it possible to find any real world objects for such a theory?
If we find such an object, does the relative phase in state vector
play any roles in such situation? And further more, the vector
space gives us the freedom to choose our base vectors, does such
transformation play any roles? The expanded strategy space
provides another class of possible state. In the mixture classical
density matrix, a system density matrix ia always has the form as
equ(\ref{systemdensity}), which is a direct product of all single
players. If pure state is permitted, a general system density
matrix may not be a direct product, but an entangled state of all
single players. Does such an entangled density matrix have
significant effect on Game Theory?

At the last of this section, lets come back to question of the
meaning of the off-diagonal term of payoff matrix $H^{i}$ by
referring to an example. What's the meaning of $\epsilon$ in the
payoff matrix below?
\begin{equation}
H^{1}=\left[\begin{array}{cccc} -2 & 0 & 0 & \epsilon_{1}
\\0 & -5 & 0 & 0
\\0 & 0 & 0 & 0
\\\epsilon_{2} & 0 & 0 & -4
\end{array}\right], H^{2}=\left[\begin{array}{cccc}
-2 & 0 & 0 & \epsilon_{2}
\\0 & 0 & 0 & 0
\\0 & 0 & -5 & 0
\\\epsilon_{1} & 0 & 0 & -4
\end{array}\right].
\end{equation}

\subsection{The new representation of static quantum game, with
quantum penny flip game as an example}
\label{qqgame}

The representation of classical game above strongly depends on the
base vectors of strategy set. But in classical game, such base
vectors are predetermined and artificial. They are just the
original discrete basic classical strategies. No inner product has
been predefined between them before our construction of the new
representation. Now we turn to Quantum Game Theory, and
fortunately it will provide us a very natural explanation of our
base vectors.

The proposed and developing Quantum Game Theory is different with
our Quantized Classical Game Theory. While our approach is a
representation, the Quantum Game Theory is a quantum version of
Game Theory. It use the idea of Game Theory, but all operations
(strategies) and the object of such operations are from quantum
world\cite{jens}. A typical 2-player quantum game is defined by
\begin{equation}
\Gamma = \left(\mathcal{H}, \rho_{0}, S_{A}, S_{B}, P_{A},
P_{B}\right), \label{qgame}
\end{equation}
in which $\mathcal{H}$ is the Hilbert space of the state of a
quantum object, $\rho_{0}$ is the initial state of such an object,
$S_{A}, S_{B}$ is player $A$ or $B$'s set of quantum operators
acting on $\mathcal{H}$. And $P_{A}, P_{B}$ are their payoff
functions.

Using well-known Quantum Penny Flip Game\cite{meyer} as an
example, spin of an electron is used as penny, so the base vectors
of $H$ are $\left|U\right>, \left|D\right>$. The initial state is
chosen as $\rho_{0} = \left|U\right>\left<U\right|$. The classical
operators are to flip the penny or not, so they are
\[
 N = \left[\begin{array}{cc}1 & 0\\0 & 1\end{array}\right], F =
\left[\begin{array}{cc}0 & 1\\1 & 0\end{array}\right].
\]
The quantum operator can be a general unitary operator
\[
\hat{U}\left(\theta,\phi\right) =
\left[\begin{array}{cc}\cos\theta & \sin\theta
e^{-i\phi}\\\sin\theta e^{i\phi} & -\cos\theta\end{array}\right].
\]
The payoff function is usually defined as
\[
E^{A} = Tr\left(P_{A}\rho_{e}\right) = -E^{B},
\]
in which
\[
P_{A} = \left[\begin{array}{cc}1& 0\\0 & -1\end{array}\right] =
-P_{B}
\]
means, player $A$ gets $1$ when the quantum object ends in
$\left|U\right>$ state and loses $1$ when in $\left|D\right>$
state.

Considering the relation between classical game and quantum game,
we redefined quantum game with a slightly difference with the
definition equ(\ref{qgame}) as
\begin{equation}
\Gamma = \left(\mathcal{H}, \rho_{0}, \left(S^{q}_{A},
S^{q}_{B}\right),  \left(S^{c}_{A}, S^{c}_{B}\right), P_{A},
P_{B}\right), \label{cgame}
\end{equation}
in which $S^{q}_{i}$ is the set of quantum operators while
$S^{c}_{i}$ is the set of classical operators, usually $S^{c}_{i}$
is a subset of $S^{q}_{i}$, but not necessary.

\subsubsection{Base vectors and strategy space}

Now let's use our new mathematical representation to reexpress the
Quantum Game Theory. Because Quantum Game Theory is constructed on
the basis of quantum state of a quantum object, $\mathcal{H}$, it
provides a set of natural base vectors of strategy space. For
quantum penny flip game, all strategy are operators with the form
of
\begin{equation}
A = A_{uu}\left|U\right>\left<U\right| +
A_{ud}\left|U\right>\left<D\right| +
A_{du}\left|D\right>\left<U\right| +
A_{dd}\left|D\right>\left<D\right|.
\end{equation}
So the base vectors are $H^{\mathcal{H}} =
\left\{\left|U\right>\left<U\right|, \left|U\right>\left<D\right|,
\left|D\right>\left<U\right|,
\left|D\right>\left<D\right|\right\}$. Furthermore, if we define
the inner product of operator as
\begin{equation}
\left<A|B\right> = Tr\left(A^{\dag}B\right),
\end{equation}
$H^{\mathcal{H}}$ is a complete orthogonal base vector set for all
quantum operators. Then the operators can be regarded as vectors
in Hilbert space, a Hilbert space of operator, which we denote as
$H^{\mathcal{H}} = \left\{\left|UU\right>, \left|UD\right>,
\left|DU\right>, \left|DD\right>\right\}$.

The classical strategies are
\begin{equation}
\left|N^{c}\right> = \left|UU\right> + \left|DD\right>,
\left|F^{c}\right> = \left|UD\right> + \left|DU\right>.
\end{equation}
In order to form another complete orthogonal base set, we need to
define other two base vectors as
\begin{equation}
\left|N^{q}\right> = \left|UU\right> - \left|DD\right>,
\left|F^{q}\right> = -i\left|UD\right> + i\left|DU\right>.
\end{equation}
Then operator space can also be expressed by $H^{\mathcal{H}} =
\left\{\left|N^{c}\right>, \left|F^{c}\right>, \left|N^{q}\right>,
\left|F^{q}\right>\right\}$, while the classical strategy is
$H^{c} =  \left\{\left|N^{c}\right>, \left|F^{c}\right>\right\}$.
$\left|UU\right>$ is the operator to turn the $\left|U\right>$
into $\left|U\right>$, no definition when the initial state is
$\left|D\right>$. A meaningful operator should give the end result
for starting state both as $\left|U\right>$ and $\left|D\right>$,
so operators $\left\{\left|N^{c}\right>, \left|F^{c}\right>,
\left|N^{q}\right>, \left|F^{q}\right>\right\}$ are better than
$\left\{\left|UU\right>, \left|UD\right>, \left|DU\right>,
\left|DD\right>\right\}$ in this. Another advantage is that all
the base vector are unitary and hermitian operator. It's easy to
prove that under our definition of inner product, the matrix form
of $\left<A\right|$ is just $A^{\dag}$. For such base strategies,
$A^{\dag} = A$, so $\left|A\right>$ and $\left<A\right|$ have the
same the matrix forms. But $\left<A\left|A\right>\right. =
Tr\left(I\right) = 2$. So a unitary operator is not in unit
magnitude. One way to solve such conflict is redefine base vectors
as $A\rightarrow\frac{1}{\sqrt{2}}A$, but here we prefer another
way to keep the form of unitary operator, and to redefined inner
product as
\begin{equation}
\left(A,B\right) = \frac{Tr\left(A^{\dag}B\right)}
{Tr\left(I\right)}.
\end{equation}

Applying our representation onto this quantum penny flip game, the
system state is
\begin{equation}
\rho^{s} = \rho^{1} \times \rho^{2}.
\label{quantumdensity}
\end{equation}
The single-player state coming from classical sub strategy space
is
\begin{equation}
\rho^{i}_{c} =
\left(p^{i}_{nc}\left|N^{c}\right>\left<N^{c}\right| +
p^{i}_{fc}\left|F^{c}\right>\left<F^{c}\right|\right).
\label{c}
\end{equation}
If we quantize it anyway as we did in section
$\S$\ref{quantizedclassical} that in a pure state of quantized
classical game, the single-player state is
\begin{equation}
\rho^{i}_{q} =
x^{i}_{nc}\bar{x}^{i}_{nc}\left|N^{c}\right>\left<N^{c}\right| +
x^{i}_{nc}\bar{x}^{i}_{fc}\left|N^{c}\right>\left<F^{c}\right| +
x^{i}_{fc}\bar{x}^{i}_{nc}\left|F^{c}\right>\left<N^{c}\right| +
x^{i}_{fc}\bar{x}^{i}_{fc}\left|F^{c}\right>\left<F^{c}\right|
\label{q}
\end{equation}
Then $\rho^{s}_{q}$ will have off-diagonal term, while
$\rho^{s}_{c}$ only has the diagonal term. Now, applying our
representation onto quantum strategy space of this penny flip
game, The single-player state is
\begin{equation}
\rho^{i}_{Q} = \sum_{\mu,\nu}
x^{i}_{\mu}\bar{x}^{i}_{\nu}\left|\mu\right>\left<\nu\right|.
\label{Q}
\end{equation}
Here $\mu,\nu$ can be any one of $\left\{\left|F^{c}\right>,
\left|N^{c}\right>, \left|F^{q}\right>,
\left|N^{q}\right>\right\}$. Next step, we allow our strategy can
be mixture state in $H^{\mathcal{H}}$. The single-player state of
a general mixture strategy will be
\begin{equation}
\rho^{i}_{Q,q} = \sum_{\mu}
p^{i}_{\mu}\left|\mu\right>\left<\mu\right|,
\label{Qq}
\end{equation}
where $\left|\mu\right>$ are quantum pure strategies, may or may
not equal to $\left\{ \left|N^{c}\right>, \left|F^{c}\right>,
\left|N^{q}\right>, \left|F^{q}\right>\right\}$. This means
$\rho^{i}_{Q,q}$ is mixture state but maybe diagonal in other set
of base vectors. A more general system state can be constructed in
the quantum strategy space $H^{\mathcal{H}}$ by destroying
equ(\ref{quantumdensity}). We ever mentioned in the last part of
section $\S$\ref{quantizedclassical} that density matrix of a
general state is not required to be a direct product to density
matrix of every single player. But still, the meaning of such
state is not clear here.

In the later discussion, we name equ(\ref{c}), equ(\ref{q}),
equ(\ref{Q}) and equ(\ref{Qq}) as classical game (CG), quantized
classical game (QCG), pure-strategy quantum game (PQG) and quantum
game (QG), respectively. And in the classical game, when the
system density matrix is not a direct product, we call the game as
entangled classical game (ECG), while for quantum case, entangled
quantum game (EQG). The strategy space of all these games have the
relation that
\begin{equation}
\left\{
\begin{array}{cc} \mbox{\bf{CG}} = \mbox{\bf{QCG}}
\subseteq \mbox{\bf{ECG}}, & \mbox{for classcial game}
\\
\mbox{\bf{PQG}} \subseteq \mbox{\bf{QG}} \subseteq
\mbox{\bf{EQG}}, & \mbox{for quantum game}
\end{array}
\right. \label{size}
\end{equation}
The relation is `$\subseteq$' not `$\subset$' because it's
possible that the later has no independent meaning other than the
former although the later has a larger strategy space. In fact,
for classical game, we still have another smaller strategy space
--- classical pure strategy space, and the game in that space ---
pure-strategy classical game (PCG). Fortunately, the relation
between PCG and CG is already clear enough through Nash Theorem,
so it's not necessary to discuss it anymore.

\subsubsection{The payoff matrix and its non-zero off-diagonal elements}
\label{spayoffmatrix}

Now all games have been unified in our mathematical
representation. Everyone has its own strategy space and base
strategy vectors. In order to finish presenting our
representation, we need to calculate the new payoff function
$H^{i}$. Let's still use the quantum penny flip game as an
example. In Quantum Game, because the non-commutative relation
between operators (base vectors), the order of acting effects the
results. On the contrary, in classical game, usually the base
vectors are commutative, so the order of acting doesn't matter. We
can see this by
\[
\left[N^{c},F^{c}\right] = 0,
\]
but
\[
\left[N^{q}, F^{q}\right] = \left[\begin{array}{cc}0 & -2i\\-2i &
0
\end{array}\right] = -2iF^{c} \neq 0.
\]
We define the order is $\left(1,2,1,2,\dots\right)$, then the
original payoff function $G^{i}$ is defined to take the value of
\begin{equation}
\left<\hat{U}^1\right|G^{i}\left|\hat{U}^2\right> =
E^{i}(\hat{U}^{1},\hat{U}^2) =
Tr\left(P^{i}\hat{U}^{2}\hat{U}^{1}\rho_{0}\left(\hat{U}^{1}\right)^{\dag}\left(\hat{U}^{2}\right)^{\dag}\right),
\label{quantumsinglepayoff}
\end{equation}
where $\hat{U}^{i}$ is anyone of $\left\{ N^{c},F^{c},
N^{q},F^{q}\right\}$. This will give all $4\times4$ values of
$\left(G^{i}_{\mu\nu}\right)_{L_{1}\times L_{2}}$.

In classical game, we require and notice that $G^{i}$ is naturally
a $\left(0,2\right)$-tensor, because the payoff of mixture
strategy is the weighted average with their own probability. The
linear property of this payoff is a requirement of our new
system-level payoff function, which is a
$\left(1,1\right)$-tensor, or we say, linear for right vector,
anti-linear for left vector. But, here in quantum game, from
equ(\ref{quantumsinglepayoff}) we find $G^{i}$ is definitely not a
tensor, not a linear mapping,
\[
G\left(\alpha \hat{U}^{1},\cdot\right) = \alpha^{*}\alpha G\left(
\hat{U}^{1},\cdot\right)\neq \alpha G\left(
\hat{U}^{1},\cdot\right).
\]
Then is it possible to transform such payoff into system-level
$\left(1,1\right)$-tensor? We need to prove it. From classical
game, one thing we already know that in the classical state
subspace $H^{c} = \left\{ N^{c},F^{c}\right\}$, no matter the
strategy is pure or mixture, such transformation exists. So we use
three steps to prove that a system-level $\left(1,1\right)$ tensor
payoff can be constructed.

First, for system state only staying on one base vectors
$\left\{N^{c}, F^{c}, N^{q}, F^{q}\right\}$, so that
\begin{equation}
\begin{array}{ccc}
\rho^{s} = \left|S\right>\left<S\right| & and & \left|S\right> =
\left|s^{1},s^{2}\right>
\end{array},
\end{equation}
We define the elements
\begin{equation}
\begin{array}{c}
H^{i}_{SS}: = \left<S\right|H^{i}\left|S\right> =
Tr\left(P^{i}s^{2}s^{1}\rho_{0}\left(s^{1}\right)^{\dag}\left(s^{2}\right)^{\dag}\right)
= G^{i}_{s^{1}s^{2}}.
\\H^{i}_{SS^{'}}: =
\left<S\left|H^{i}\right|S^{'}\right> =
Tr\left(P^{i}s^{2'}s^{1'}\rho_{0}\left(s^{1}\right)^{\dag}\left(s^{2}\right)^{\dag}\right)
\end{array}.
\label{quantumpayoffmatrix}
\end{equation}
For example, $H^{1}_{nc,nc;nc,nc} = 1$, which means when both
player $1$ and player $2$ choose $F^{c}$, player $1$ wins;
$H^{1}_{nc,nc;nq,nc} = 1$, which has no classical meaning, because
it's a off-diagonal elements. It's easy to prove that $H^{i}$ is
hermitian,
\[
\begin{array}{ccc}
H^{i}_{S^{'}S} & = & \left<S^{'}\left|H^{i}\right|S\right>
\\ & = &
Tr\left(P^{i}s^{2}s^{1}\rho_{0}\left(s^{1'}\right)^{\dag}\left(s^{2'}\right)^{\dag}\right)
\\ & = & \left[Tr\left(P^{i}s^{2}s^{1}\rho_{0}\left(s^{1'}\right)^{\dag}\left(s^{2'}\right)^{\dag}\right)^{\dag}\right]^{*}
\\ & = & \left[Tr\left(s^{2'}s^{1'}\rho^{\dag}_{0}s^{1}s^{2}\left(P^{i}\right)^{\dag}\right)\right]^{*}
\\ & = & \left[Tr\left(P^{i}s^{2'}s^{1'}\rho^{\dag}_{0}s^{1}s^{2}\right)\right]^{*}
\\ & = & \left(H^{i}_{SS^{'}}\right)^{*},
\end{array}
\]
in which, we require $\left(P^{i}\right)^{\dag} = P^{i}$. So
\begin{equation}
\left(H^{i}\right)^{\dag} = H^{i}.
\end{equation}

Second, we prove that for a system state not staying on the base
vector, but on pure state, such as
\[
\begin{array}{ccc}
\rho^{s} = \left|S\right>\left<S\right| & and & \left|S\right> =
x^{1}_{1}\left|s^{1}_{1},s^{2}\right> +
x^{1}_{2}\left|s^{1}_{2},s^{2}\right>
\end{array},
\]
we still have $E^{i}\left(S\right) = Tr\left(\rho^{s}H^{i}\right)
= \left<S\right|H^{i}\left|S\right>$, in which $H^{i}$ is a
$(1,1)$-tensor.

\noindent{\bf{Proof}}: from payoff definition
equ(\ref{quantumsinglepayoff}),
\[
\begin{array}{lll}
E^{i}(x^{1}_{1}\left|s^{1}_{1}\right> +
x^{1}_{2}\left|s^{1}_{2}\right>,\left|s^{2}\right>)  & = &
Tr\left(P^{i}s^{2}\left(x^{1}_{1}s^{1}_{1}+ x^{1}_{2}s^{1}_{2}
\right)\rho_{0}\left(x^{1}_{1}s^{1}_{1}+ x^{1}_{2}s^{1}_{2}
\right)^{\dag}\left(s^{2}\right)^{\dag}\right)
\\ & = & x^{1}_{1}\bar{x}^{1}_{1}Tr\left(P^{i}s^{2}s^{1}_{1}\rho_{0}\left(s^{1}_{1}
\right)^{\dag}\left(s^{2}\right)^{\dag}\right) +
\\ & & x^{1}_{1}\bar{x}^{1}_{2}Tr\left(P^{i}s^{2}s^{1}_{1}\rho_{0}\left(s^{1}_{2}
\right)^{\dag}\left(s^{2}\right)^{\dag}\right) +
\\ & & x^{1}_{2}\bar{x}^{1}_{1}Tr\left(P^{i}s^{2}s^{1}_{2}\rho_{0}\left(s^{1}_{1}
\right)^{\dag}\left(s^{2}\right)^{\dag}\right) +
\\ & & x^{1}_{2}\bar{x}^{1}_{2}Tr\left(P^{i}s^{2}s^{1}_{2}\rho_{0}\left(s^{1}_{2}
\right)^{\dag}\left(s^{2}\right)^{\dag}\right),
\end{array}
\]
in which we need the property that $Tr\left(\cdot\right)$ is a
linear operator. On the other hand, when $H^{i}$ is a
$\left(1,1\right)$-tensor,
\[
\begin{array}{lll}
\left<S\right|H^{i}\left|S\right>
 & = & \left(\bar{x}^{1}_{1}\left<s^{1}_{1},s^{2}\right| +
\bar{x}^{1}_{2}\left<s^{1}_{2},s^{2}\right|\right)
H^{i}\left(x^{1}_{1}\left|s^{1}_{1},s^{2}\right> +
x^{1}_{2}\left|s^{1}_{2},s^{2}\right>\right)
\\ & = & x^{1}_{1}\bar{x}^{1}_{1}\left<s^{1}_{1},s^{2}\right|H^{i}\left|s^{1}_{1},s^{2}\right> +
\\ &   & x^{1}_{1}\bar{x}^{1}_{2}\left<s^{1}_{2},s^{2}\right|H^{i}\left|s^{1}_{1},s^{2}\right> +
\\ &   & x^{1}_{2}\bar{x}^{1}_{1}\left<s^{1}_{1},s^{2}\right|H^{i}\left|s^{1}_{2},s^{2}\right> +
\\ &   & x^{1}_{2}\bar{x}^{1}_{2}\left<s^{1}_{2},s^{2}\right|H^{i}\left|s^{1}_{2},s^{2}\right>
\end{array}
\]
Using the definition of $H^{i}_{SS^{'}}$ in
equ(\ref{quantumpayoffmatrix}), we know they equal.

At last, we prove for a mixture state, such as
\[
\rho^{s} =
p^{1}_{1}\left|s^{1}_{1},s^{2}\right>\left<s^{1}_{1},s^{2}\right|
+
p^{1}_{2}\left|s^{1}_{2},s^{2}\right>\left<s^{1}_{2},s^{2}\right|,
\]
we still have $E^{i}\left(S\right)=Tr\left(\rho^{s}H^{i}\right)$.

\noindent{\bf{Proof}}: using above result,
\[
\begin{array}{lll}
Tr\left(\rho^{s}H^{i}\right) & = &
\sum_{\mu,\nu}\left<\mu,\nu\left|\rho^{s}H^{i}\right|\mu,\nu\right>
\\ & = &
p^{1}_{1}\left<s^{1}_{1},s^{2}\left|H^{i}\right|s^{1}_{1},s^{2}\right>
+
p^{1}_{2}\left<s^{1}_{2},s^{2}\left|H^{i}\right|s^{1}_{2},s^{2}\right>
\\& = & p^{1}_{1}E^{i}\left(s^{1}_{1},s^{2}\right) + p^{1}_{2}E^{i}\left(s^{1}_{2},s^{2}\right)
\\& = & E^{i}\left(S\right)
\end{array}
\]
Therefor, for any system state we still have
\begin{equation}
E^{i}\left(S\right) = Tr\left(\rho^{s}H^{i}\right).
\label{payoffvalue}
\end{equation}
The payoff matrix of the quantum penny flip game is a $16\times16$
matrix
\[
H^{1} = \left[\begin{array}{cccccccccccccccc}
1&0&1&0&0&1&0&i&1&0&1&0&0&-i&0&1
\\0&-1&0&i&-1&0&1&0&0&-1&0&i&i&0&-i&0
\\1&0&1&0&0&1&0&i&1&0&1&0&0&-i&0&1
\\0&-i&0&-1&-i&0&i&0&0&-i&0&-1&-1&0&1&0
\\0&-1&0&i&-1&0&1&0&0&-1&0&i&i&0&-i&0
\\1&0&1&0&0&1&0&i&1&0&1&0&0&-i&0&1
\\0&1&0&-i&1&0&-1&0&0&1&0&-i&-i&0&i&0
\\-i&0&-i&0&0&-i&0&1&-i&0&-i&0&0&-1&0&-i
\\1&0&1&0&0&1&0&i&1&0&1&0&0&-i&0&1
\\0&-1&0&i&-1&0&1&0&0&-1&0&i&i&0&-i&0
\\1&0&1&0&0&1&0&i&1&0&1&0&0&-i&0&1
\\0&-i&0&-1&-i&0&i&0&0&-i&0&-1&-1&0&1&0
\\0&-i&0&-1&-i&0&i&0&0&-i&0&-1&-1&0&1&0
\\i&0&i&0&0&i&0&-1&i&0&i&0&0&1&0&i
\\0&i&0&1&i&0&-i&0&0&i&0&1&1&0&-1&0
\\1&0&1&0&0&1&0&i&1&0&1&0&0&-i&0&1
\end{array}
\right]
\]
and $H^{2} = -H^{1}$. Compare with the payoff matrix of classical
game with the payoff matrix of quantum game, a significant
difference is that the later has non-zero off-diagonal elements
while the former only has diagonal elements. Through this
representation we know the difference between classical game and
quantum game is not only the size of strategy space but also the
off-diagonal elements of payoff matrix.

From $H^{i}$ above, the sub-matrix related with $\left\{N^{c},
F^{c}\right\}$ is
\[
H^{1,c} = \left[\begin{array}{cccc}1 & 0 & 0 & 1\\
0 & -1 & -1 & 0\\
0 & -1 & -1 & 0\\
1 & 0 & 0 & 1
\end{array}
\right]= -H^{2,c}.
\]
They are different with the new payoff matrix of the original
classical game,
\[
H^{i,c}_{original}\left[\begin{array}{cccc}1 & 0 & 0 & 0\\
0 & -1 & 0 & 0\\
0 & 0& -1 & 0\\
0 & 0 & 0 & 1
\end{array}
\right]= -H^{2,c}_{original},
\]
which only has diagonal terms. So we can say, the quantization
process changes the definition of the original classical game.
However, the meaning of quantum game does not completely depends
on the equivalence with the original classical game. So even given
such difference, quantum game is still probably a new game.

If we defined a quantum game by payoff matrix $H^{i}$, another
privilege of this new representation is that the definition of a
quantum game is independent on $\left(\mathcal{H}, \rho\right)$,
the state of a quantum object. Our payoff function can be directly
defined on system state $\rho^{S}$. Of course, any payoff defined
on $\left(\mathcal{H}, \rho\right)$ can be transferred
equivalently into a function on $\rho^{S}$. So a quantum game can
be defined as
\begin{equation}
\Gamma = \left(\prod_{i}^{N}\left(\times S^{i,q}\right),
\prod_{i}^{N}\left(\times S^{i,c}\right),
\left\{H^{i}\right\}\right).
\label{ourquantumgame}
\end{equation}
in which $S^{i,q}$ has base vectors
$\left\{\left|s^{i,q}_{\mu}\right>\right\}$, and $S^{c}_{i}$ has
base vectors $\left\{\left|s^{i,c}_{\nu}\right>\right\}$. Usually
the later is a subset of the former. A classical payoff function
is defined on system base vectors such as
$H^{i,c}=\sum_{S}\left|S\right> H^{i,c}_{SS}\left<S\right|$, while
a quantum payoff function is defined as $H^{i} = \sum_{SS^{'}}
\left|S\left> H^{i}_{SS^{'}}\right<S^{'}\right|$. If $S^{c}_{i}$
is a subspace of $S^{q}_{i}$, the payoff matrix of classical game
just take the corresponding diagonal parts whether the
corresponding sub-matrix of the quantum game payoff matrix has
non-zero off-diagonal terms or not.

In the next part of this paper, we try to make use of this
representation and hopefully to find something significant.

\section{Pseudo-Dynamical Theory of Equilibrium State}
\label{statphys}

In the new representation, a game seems very similar with an Ising
model with global interaction. The payoff of every player is
related with everyone else. The state of every player can be
represented by a quantum state vector or density matrix. Every
player try to stay at point with the maximum payoff, while in
Ising model, the whole system try to stay at minimum energy point.
The distribution of a quantum system at thermal equilibrium is
\begin{equation}
\begin{array}{ccc}
\rho = \frac{1}{Z}e^{-\beta H} & and & Z = Tr\left(e^{-\beta
H}\right),
\end{array}
\end{equation}
where $H$ is the system Hamiltonian, $Z$ is so called partition
function.

Now as in Statistical Mechanics, we introduce the idea of
distribution function of state into Game Theory, but instead of
function in $\Gamma$ space in Statistical Mechanics, here in $\mu$
space, the state space of every single player. A natural form is
\begin{equation}
\begin{array}{ccc}
\rho^{i} = \frac{1}{Z}e^{\beta H^{i}_{R}} & and & Z =
Tr\left(e^{\beta H^{i}_{R}}\right),
\end{array}
\end{equation}
in which $H^{i}_{R}$ is the payoff function of player $i$ in its
own strategy space and $Z$ is the partition function in $i$'s
strategy space. The payoff matrix $H^{i}$ we have now is defined
in system strategy space. So a kind of reduced matrix is what we
need to find.

Before the detailed calculation, one thing we should notice that
the equilibrium density matrix description is different with the
classical mixture strategy. If the eigenvectors of $H^{i}_{R}$ can
be found as $\left\{\left|\mu\right>^{\epsilon}\right\}$, then
\begin{equation}
\rho^{i} =
\sum_{\mu}p^{i}_{\mu,\epsilon}\left|\mu\right>^{\epsilon}\left<\mu\right|^{\epsilon}
\end{equation}
is similar with the classical mixture strategy form, and
$p^{i}_{\mu,\epsilon}$ can be regarded as the probability on
strategy $\mu^{\epsilon}$. But first, such an set of eigenvectors
is not always the same as the classical base vectors, because
sometimes, we have non-zero off-diagonal elements. Second, such a
density matrix gives the probability of any pure strategies
$\left|s\right>$ even being different with the base vector, by
\begin{equation}
p^{i}_{s} = \left<s\left|\rho^{i}\right|s\right>.
\end{equation}
This is impossible in mixture strategy description.

\subsection{Reduced payoff matrix and Kinetics Equation for
Equilibrium State}

Now we start to define the reduced payoff matrix and investigate
its properties. A Nash Equilibrium state is defined that at that
point every player is at the maximum point due to the choices of
all other players are fixed. A reduced payoff matrix should
describe the payoff of a single person when the choice of all
other players. In the traditional language of Game Theory, such a
reduced payoff matrix is equivalently to be defined like the end
result of equ(\ref{oldearn}) under any arbitrary fixed
$\left|P^{2}\right>^{old}$. But we need a matrix form here.

For pure system strategy,
$H^{i}\left(\left<S\right|;\left|S\right>\right)$ is a
$\left(1,1\right)$-tensor. In a 2-player game,
$H^{i}\left(\left<s^{1}\right|,\left<s^{2}\right|;\left|s^{1}\right>,\left|s^{2}\right>\right)$
can also be regarded as a $\left(2,2\right)$-tensor. A reduced
payoff matrix of player $i$ means in the viewpoint of player $i$
it should be a $\left(1,1\right)$-tensor. When both player $1$ and
player $2$ stays on pure strategy $\left|s^{1}\right>,
\left|s^{2}_{fixed}\right>$ respectively, it has a natural
definition,
$H^{1}_{R}\left(\left<s^{1}\right|;\left|s^{1}\right>\right) =
H^{1}\left(\left<s^{1}\right|,\left<s^{2}_{fixed}\right|;\left|s^{1}\right>,\left|s^{2}_{fixed}\right>\right)$.
Since our strategy can be a mixture state, or generally a density
matrix form, we need to generalize the above definition. A reduced
payoff matrix of player $1$ in a $2$-player game is defined as
\begin{equation}
H_{R}^{1} = Tr^{2}(\rho^{2}_{fixed}H^{1}),
\label{reducedpayoff}
\end{equation}
where $Tr^{2}$ is the trace in subsapce of player $2$. From
equ(\ref{payoffvalue}), the payoff value of player $1$ is
\[
\begin{array}{lll}
E^{1} & = & Tr\left(\rho^{1}\rho^{2}H^{1}\right)
\\    & = &
\sum_{S}\left<S\right|\rho^{1}\rho^{2}H^{1}\left|S\right>
\\    & = & \sum_{\gamma\nu}\sum_{\alpha\beta}\left<\gamma\nu\right|\rho^{1}_{\alpha\beta}\left|\alpha\right>\left<\beta\right|\rho^{2}H^{1}\left|\gamma\nu\right>
\\    & = & \sum_{\alpha\beta}\rho^{1}_{\alpha\beta}\sum_{\nu}\left<\nu\right|\left<\beta\right|\rho^{2}H^{1}\left|\alpha\right>\left|\nu\right>
\\    & = & \sum_{\alpha\beta}\rho^{1}_{\alpha\beta}\left<\beta\right|Tr^{2}\left(\rho^{2}H^{1}\right)\left|\alpha\right>
\\    & = & \sum_{\alpha\beta}\rho^{1}_{\alpha\beta}\left<\beta\right|H^{1}_{R}\left|\alpha\right>
\\    & = & Tr^{1}\left(\rho^{1}H^{1}_{R}\right)
\end{array}
\]
So if we know the reduced payoff matrix of player $1$, the payoff
value can be calculated by
\begin{equation}
E^{1} = Tr^{1}(\rho^{1}H^{1}_{R}).
\end{equation}

In fact the $Tr^{2}$ action is quite hard to perform, because this
requires the result of a trace is a matrix, not a number as usual.
An equivalent but easily understood form of
equ(\ref{reducedpayoff}) is
\[
\left(H_{R}^{1}\right)_{\alpha\beta} =
Tr^{2}(\rho^{2}_{fixed}H_{\alpha\beta}^{1}),
\]
in which $H^{1}_{\alpha\beta}$ is a sub matrix with fixed player
$1$'s index (here, first and third index). In order to define a
general form for $N$-player game, we denote the trace $Tr_{-i}$ as
diagonal summation in the space except player $i$'s. So in
$2$-player game, $Tr^{1} = Tr_{2}$. Then a general reduced payoff
matrix of player $i$ under fixed strategies of all other players
is
\begin{equation}
H_{R}^{i} =
Tr_{-i}(\rho^{1}\cdots\rho^{i-1}\rho^{i+1}\cdots\rho^{N}H^{i}).
\end{equation}

Still using the Prisoner's Dilemma as example, when player $2$
choose strategy $C$ with $p^{2}_{c}$ and $D$ with $p^{2}_d$, the
state is
\[
\rho^{2}=\left[\begin{array}{cc}p^{2}_{c} & 0
\\0 & p^{2}_{d}
\end{array}
\right]
\]
Then
\[
\begin{array}{lll}
H_{R}^{1} & = & Tr_{-1}\left(\left[\begin{array}{cc}p^{2}_{c} & 0
\\0 & p^{2}_{d}
\end{array}
\right]\left[\begin{array}{cc|cc} -2 & 0 & 0 & 0
\\0 & -5 & 0 & 0
\\ \hline 0 & 0 & 0 & 0
\\0 & 0 & 0 & -4
\end{array}\right]\right)
\\ & = & \left[
\begin{array}{cc}
-2p^{2}_{c}-5p^{2}_{d} & 0
\\0 & 0\cdot p^{2}_{c}-4p^{2}_d
\end{array}\right].
\end{array}
\]

Recalls Metropolis Method and its derivative Heat
Bath\cite{heatbath} Method in Monte Carlo Simulation of
Statistical Ensemble. In the simulation of equilibrium state of
Ising model, every single step, when a random spin is chosen, it
faces the same situation with our game player. All other spins
have decided one state to stay temporary, it has some choices of
its own state by evaluating the energy difference between all its
possible states. Then it choose one state to stay by a transition
probability or transition rate over all possible states. The
Kinetics Equation for such process is not unique, different forms
of transition probability can give the same equilibrium state.

Now we face a quantum system, although a similar situation. Every
player should make his decision every step with the fixed state of
all other players and we also ask for the equilibrium state. The
reason that different Kinetics Equations give the same equilibrium
state in Statistical Physics is the well-known Detailed Balanced
Theorem in thermal equilibrium, but we don't have a corresponding
one in Game Theory. We now just suppose that at equilibrium state,
the density matrix of player $i$'s state is
\begin{equation}
\begin{array}{ccc}
\rho^{i} = \frac{1}{Z}e^{\beta H^{i}_{R}} & and & Z =
Tr^{i}\left(e^{\beta H^{i}_{R}}\right).
\end{array}
\label{equilibriumdistribution}
\end{equation}
And we choose a heuristic Kinetics Equation as iteration equation,
\begin{equation}
\rho^{i}\left(t\right) = \frac{1}{Z\left(t-1\right)}e^{\beta
H^{i}_{R}\left(t-1\right)}
\label{kineticsequation}
\end{equation}
Then the equilibrium state is defined as the fixed point of this
iteration if it has fixed point.

\subsection{Examples and the effect of $\beta$}
\label{phase}

In fact, Kinetics Equation equ(\ref{kineticsequation}) is $N$
related iteration equations. The existence of the fixed point is
not obvious. Even the questions itself is not unique, although the
experience in simulation in Statistical Physics implies that such
equation should exist probably with different form. The fixed
point might be different with Nash Equilibrium even if it exists.
In this paper, all these questions are neglected. Let's first test
such idea in some examples, just like what a physicist usually
does, not a mathematician, who will pay more attention on a
general definition of equilibrium state and the proof of the
existence.

Equ(\ref{kineticsequation}) of a classical game is much easier to
deal with than the one of a quantum game. In classical game, both
$H^{i}$ and $H^{i}_{R}$ are diagonal. The density matrix at time
$t$ can always be written as $\rho^{i}\left(t\right) =
\sum_{\alpha}p^{i}_{\alpha}\left(t\right)\left|\alpha\right>\left<\alpha\right|$,
then equ(\ref{kineticsequation}) will lead to a series of
evolution equations for $p^{i}_{\alpha}\left(t\right)$.

However, in quantum game, since the payoff matrix $H^{i}$ has
off-diagonal elements, the reduced payoff matrix $H^{i}_{R}$ also
can have off-diagonal elements. Then the density matrix can be
equivalently replaced by evolution equation of
$p^{i}_{s}\left(t\right)$ {\bf{only}} when the density matrix is
expressed in the base vector formed by the eigenvectors of
$H^{i}_{R}$. But with off-diagonal elements, such eigenvectors are
not always the base vector we used to express the game and they
might change during the iteration process. So the first step is to
solve the eigenvalue equation of $H^{i}$ and $H^{i}_{R}$.

\subsubsection{Eigenvalue Problem}
\label{eigenvalue}

The eigenvalue problem in classical game is quite easy. All the
eigenvectors are the base vector we used, the eigenvalues are just
the corresponding diagonal elements. In a quantum game, it depends
on the details of payoff matrix. For example, in the quantum penny
flip game, the payoff matrix $H^{1}$ has $16\times16$ elements.
Even when player $2$ choose $\rho^{2,fixed} =
\left|N^{c}\right>\left<N^{c}\right|$, the reduced payoff matrix
of player $1$ is
\[
H^{1}_{R} = \left[
\begin{array}{cccc}
1&0&1&0
\\0&-1&0&i
\\1&0&1&0
\\0&-i&0&-1
\end{array}
\right]
\]
The eigenvalues and the corresponding eigenvectors are
\[
\left(
\begin{array}{lll}-2 & \rightarrow & \left[0,-i,0,1\right]^{T}
\\ 2 & \rightarrow & \left[1,0,1,0\right]^{T}
\\ 0 & \rightarrow & \left[-1,0,1,0\right]^{T}
\\ 0 & \rightarrow & \left[0,i,0,1\right]^{T}
\end{array}
\right).
\]
This means a quantum player can make money over the classical
player with $\rho^{2,fixed}$ by using strategy
$\frac{\sqrt{2}}{2}\left[1,0,1,0\right]^{T}$. And the funny thing
is the value of payoff the 2, not $1$ in classical case when
player $1$ uses $\left[1,0,0,0\right]^{T}$. It clearly shows the
effect of the off-diagonal elements for a quantum player. If the
player $1$ is still a classical player, the strategy he can use is
just $N^{c}, F^{c}$, so he will get $1, -1$ respectively. Anyway,
the topic of this section is show the way to do the iteration
defined in the Kinetics Equation equ(\ref{kineticsequation}), not
the reason of such difference.

Now we have the idea. Starting, for instance, from player $2$
choose $\rho^{2}\left(t=0\right) = \rho^{2,fixed}$, the state of
player $1$ is then
\[
\rho^{1}\left(t=1\right) =
\frac{1}{e^{-2\beta}+e^{2\beta}+2e^{0\beta}}\left(e^{-2\beta
}\left|-2\right>\left<-2\right| + e^{2\beta
}\left|2\right>\left<2\right| + e^{0\beta
}\left|0_{1}\right>\left<0_{1}\right| + e^{0\beta
}\left|0_{2}\right>\left<0_{2}\right|\right).
\]
And then substitute it back to equ(\ref{kineticsequation}) and do
the iteration. However, from above density matrix we know that
even beginning from a pure state, the state after one iteration
will be a mixture state. In classical game, it doesn't matter,
because the end state generally can be a mixture state, and a pure
state is equivalent with a mixture state with the same diagonal
part. But in quantum game, mixture strategy is quite different
with a pure one. One way to deal with this problem is to set
$\beta = \infty$. Then the Kinetics Equation of quantum game
becomes,
\begin{equation}
\rho^{i}\left(t\right) =
\left|s^{i}_{Max}\right>\left<s^{i}_{Max}\right|,
\end{equation}
in which $\left|s^{i}_{Max}\right>$ is the eigenvector with
maximum eigenvalue of the reduced payoff matrix. So only the
maximum one is kept after every step. But this will brings new
problems when the real equilibrium state is a mixture state. So
for quantum game, it's better to regard the approach shown here
just as an idea. Later on, classical game is our main object. The
task of this section is just to point out that a quantum game
brings new things such as eigenvalue problem while which is quite
trivial in classical game.

\subsubsection{Equilibrium state calculation of several examples}
\begin{figure}
\includegraphics{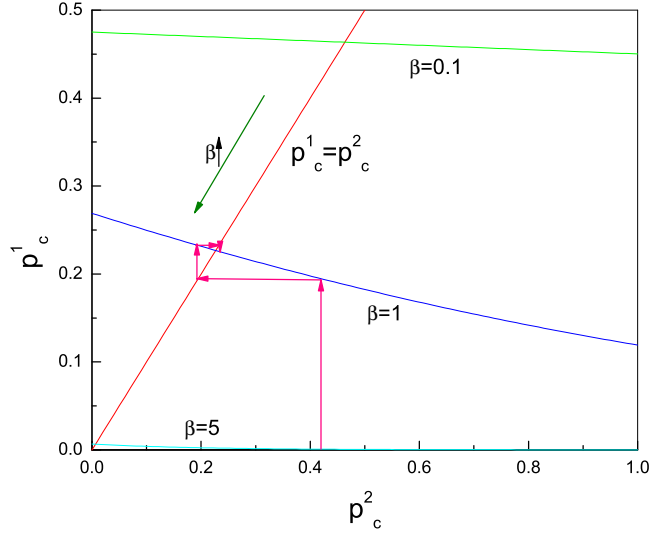}
\caption{The iteration process defined by the Kinetics Equation of
Prisoner's Dilemma drives the fixed point from a finite number to
0 when $\beta$ growths. Because here only one parameter $p_{c}$ we
need to calculate, a simple fixed point graph shows the result.
Usually in a multi-strategy game or with more players, we will
have more parameters and more complex equations. Then in that
situation, we will have to use simulation. The function plotted
here is $p_{c}=\frac{1}{1+e^{\beta\left(1+p_{c}\right)}}$. Because
the two steps of one iteration use the same function form, it can
be regarded as two iteration steps with only one function.}
\label{graphprisoner}
\end{figure}

Now we come to use our Kinetics Equation approach on some examples
of classical game. First, let's finish the discussion about the
Prisoner's Dilemma. We already know
\[
\begin{array}{ccc}
\begin{array}{lll}
H_{R}^{1} & = & \left[
\begin{array}{cc}
-2p^{2}_{c}-5p^{2}_{d} & 0
\\0 & 0\cdot p^{2}_{c}-4p^{2}_d
\end{array}\right]
\end{array}
& \mbox{and} &
\begin{array}{lll}
H_{R}^{2} & = & \left[
\begin{array}{cc}
-2p^{1}_{c}-5p^{1}_{d} & 0
\\0 &0\cdot p^{1}_{c}-4p^{1}_d
\end{array}\right].
\end{array}
\end{array}
\]
Suppose we start from player $2$ with
$\left(p^{2}_{c}\left(0\right),1-p^{2}_{c}\left(0\right)\right)$,
then from equ(\ref{kineticsequation}),
\[
\left\{
\begin{array}{l}
p^{1}_{c}\left(t\right) =
\frac{e^{\beta\left(-2p^{2}_{c}\left(t-1\right)-5p^{2}_{d}\left(t-1\right)\right)}}
{e^{\beta\left(-2p^{2}_{c}\left(t-1\right)-5p^{2}_{d}\left(t-1\right)\right)}
+ e^{\beta\left(0\cdot
p^{2}_{c}\left(t-1\right)-4p^{2}_{d}\left(t-1\right)\right)}} =
\frac{1} {1 + e^{\beta\left(1+p^{2}_{c}\left(t-1\right)\right)}}
\\
\\
p^{2}_{c}\left(t+1\right) = \frac{1} {1 +
e^{\beta\left(1+p^{1}_{c}\left(t\right)\right)}}
\end{array}
\right.
\]
When $\beta = \infty$, which means infinite resolution level, or
we say, any difference in payoff is significant, then $p^{1}_{c} =
0 = p^{2}_{c}$. The equilibrium state is $\left(D, D\right)$,
which is equivalent with Nash Equilibrium. When $\beta$ is finite,
denote the fixed point as $\left(p^{1,*}_{c},p^{2,*}_{c}\right)$.
The stability of this fixed point can be analyzed by the linear
stability matrix,
\begin{equation}
S = \left[\begin{array}{cc}\frac{\partial p^{1}_{c}}{\partial
p^{1}_{c}} & \frac{\partial p^{1}_{c}}{\partial p^{2}_{c}}
\\\frac{\partial p^{2}_{c}}{\partial
p^{1}_{c}} & \frac{\partial p^{2}_{c}}{\partial p^{2}_{c}}
\end{array}
\right]_{\left(p^{1,*}_{c},p^{2,*}_{c}\right)} \triangleq
\left(\frac{\partial p^{i}_{\alpha}}{\partial
p^{j}_{\mu}}\right)_{\left(\prod_{i}L_{i}\right)\times\left(\prod_{i}L_{i}\right)}\left|_{\left(p^{i,*}_{\nu}\right)}\right.
\end{equation}
In this specific case, it's unstable, the fixed point graph of the
Kinetics Equation is shown in fig(\ref{graphprisoner}). When
$\beta = 0$, which means the players care nothing about the
payoff, then $p^{1}_{c} = \frac{1}{2} = p^{2}_{c}$. Of course,
such solution is useless, but still consistent with our intuitive
result.

\begin{figure}
\includegraphics{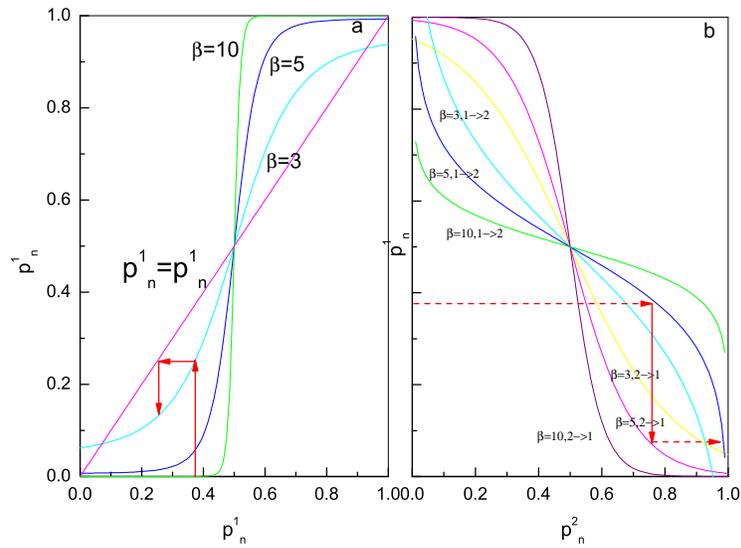}
\caption{In figure (a), the function plotted is $p_{h} =
\frac{1}{1+e^{\beta\left(2\frac{1}{1+e^{\beta\left(2p_{h}-1\right)}}-1\right)}}$.
There are three fixed points for this iteration process. One is
unstable fixed point, $0.5$, while the other two are stable fixed
points closing to $0$ and $1$ respectively, when $\beta$ grows. In
figure (b), we explicitly show the two iterations instead of the
self-mapping from $p^{1}_{h}$ to $p^{1}_{h}$ in figure (a). It
gives more details. $\left(0.5, 0.5 \right)$ is still an unstable
fixed point. While the end state can be a $\left(0,1\right)$ or
$\left(1,0\right)$ depending on the initial state.}
\label{graphhawk}
\end{figure}

Second example, we choose Hawk-Dove, a two-pure-NE game. The
payoff matrix of player $1$ and $2$ are
\[
H^{1} = \left[
\begin{array}{cccc}3 & 0 & 0 & 0
\\0 & 1 & 0 & 0
\\0 & 0 & 4 & 0
\\0 & 0 & 0 & 0
\end{array}
\right], H^{2} =\left[
\begin{array}{cccc}3 & 0 & 0 & 0
\\0 & 4 & 0 & 0
\\0 & 0 & 1 & 0
\\0 & 0 & 0 & 0
\end{array}
\right]
\]
The reduced payoff matrix is
\[
H^{i}_{R} = \left[
\begin{array}{cc}3p^{\left(3-i\right)}_{h}+1p^{\left(3-i\right)}_{d} & 0
\\ 0 & 4p^{\left(3-i\right)}_{h}
\end{array}
\right]
\]
Then the Kinetics Equation is
\[
p^{i}_{h} =
\frac{1}{1+e^{\beta\left(p^{\left(3-i\right)}_{h}-p^{\left(3-i\right)}_{d}\right)}}
\]
It's easy to know that when $\beta = \infty$, fixed point are
$\left(p^{1}_{h} = 0, p^{2}_{h} = 1\right)$, $\left(p^{1}_{h} =
\frac{1}{2}, p^{2}_{h} = \frac{1}{2}\right)$ and $\left(p^{1}_{h}
= 1, p^{2}_{h} = 0\right)$ depending the initial state
$p^{2}_{h}>p^{2}_{d}$, $p^{2}_{h}=p^{2}_{d}$ or
$p^{2}_{h}<p^{2}_{d}$. For a finite $\beta$, a fixed point graph
is shown in fig(\ref{graphhawk}).

\begin{figure}
\includegraphics{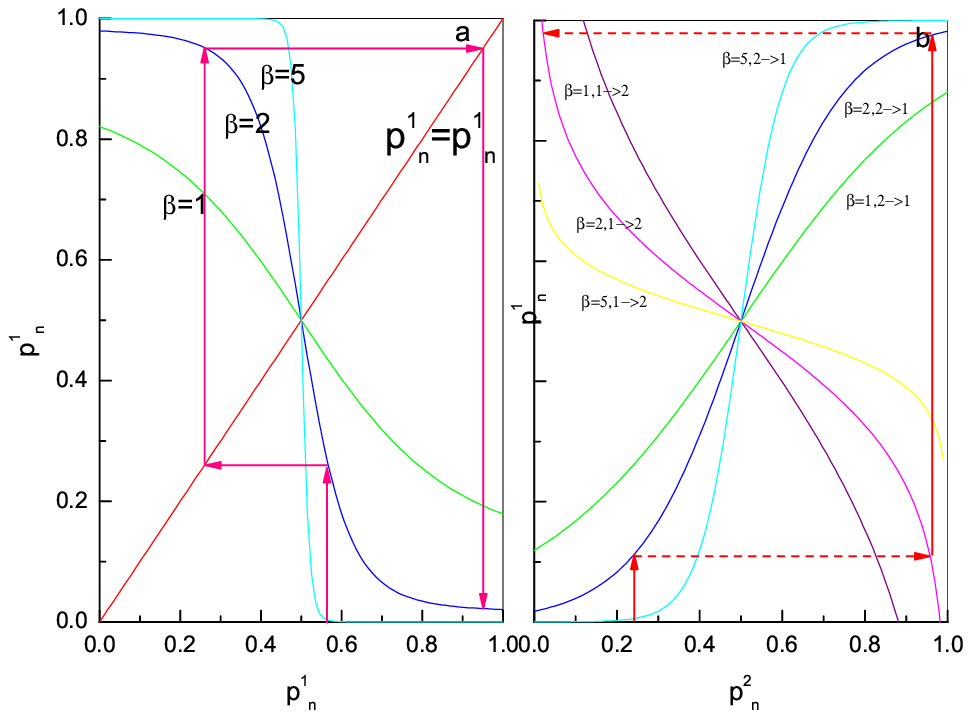}
\caption{In figure (a), the function plotted here is $p^{1}_{n} =
\frac{1}{1+e^{2\beta\left(1-2\frac{1}{1+2\beta\left(2p^{1}_{n}-1\right)}\right)}}$.
The only fixed but unstable point is always $0.5$ for different
$\beta$. In figure (b), two iteration functions are used to show a
clearer but more complex picture, from which we know
$\left(0.5,0.5\right)$ is still a fixed point, but if starts from
initial state other than this point, the system will jump between
$\left(0,0\right)$, $\left(0,1\right)$, $\left(1,1\right)$ and
$\left(1,0\right)$. } \label{graphpenny}
\end{figure}

The third example is the classical penny flip game, which has no
pure NE. The payoff matrix are
\[
H^{1} = \left[
\begin{array}{cccc}1 & 0 & 0 & 0
\\0 & -1 & 0 & 0
\\0 & 0 & -1 & 0
\\0 & 0 & 0 & 1
\end{array}
\right] = -H^{2}.
\]
The reduced payoff matrix is
\[
H^{1}_{R} = \left[
\begin{array}{cc}p^{2}_{n}-p^{2}_{f} & 0
\\ 0 & p^{2}_{f}-p^{2}_{n}
\end{array}
\right], H^{2}_{R} = \left[
\begin{array}{cc}p^{1}_{f}-p^{1}_{n} & 0
\\ 0 & p^{1}_{n}-p^{1}_{f}
\end{array}
\right].
\]
Then the Kinetics Equation is
\[
p^{1}_{n} =
\frac{1}{1+e^{2\beta\left(p^{2}_{f}-p^{2}_{n}\right)}}, p^{2}_{n}
= \frac{1}{1+e^{2\beta\left(p^{1}_{n}-p^{1}_{f}\right)}}.
\]
From fig(\ref{graphpenny}), $\left(p^{1}_{n} =
\frac{1}{2},p^{2}_{n} = \frac{1}{2}\right)$ is the only fixed
point no matter $\beta=\infty$ or not. And even the only fixed
point is unstable.

The Kinetics Equation, its fixed points and the stability analysis
of the fixed points gives a method to find equilibrium state and
to refine them if we require an applicable equilibrium point is a
stable fixed point. From all the trivial cases we tested it seems
valid. But questions such as more tests, a general form of such
equation, and the relation between such fixed points and Nash
Equilibrium is waiting for more detailed discussion.

At last, we have to admit that our simulation is not equivalent
with the Kinetics Equation. Pure strategies are included by the
Kinetics Equation, but since our algorithm is classical, here we
only let it evolute in the subspace of mixture strategy. For
classical game, this is not a fatal problem, because we have prove
that pure strategy is equivalent with mixture strategy having the
same diagonal part. But for a quantum game, pure strategy is
totaly different with the mixture classical one because of the
off-diagonal elements of payoff matrix. Is it possible to find
such a simulation algorithm?

On the other hand, when $\beta\neq\infty$, the fixed point of our
Kinetics Equation might not equal to the Nash Equilibrium state.
Such fixed points are the end states when the average resolution
level of all players is $\beta$, which can be regarded as a
typical scale which players care. This concepts may expand the
description of Game Theory into the situation that players are not
complete rational. They can evaluate the payoff, but not
explicitly, only a rough range. And from the experience in
Statistical Physics, especially Phase Transition, we know that
even when $\beta$ is not very large but large enough the lowest
energy mode (here, maximum payoff mode) will dominate the system.
This means, under some not extremely restricted conditions, the
traditional Nash Equilibrium is still valid. It will be funny if
one can prove such conclusion from a general situation in our
equilibrium definition.

\section{Discussion}
\label{discussion}

It's quite straightforward to extend our notation into $N$-player
game and continuous strategy case. However, although a new
representation has been introduced to express everything in a
static game, the advantage of such a language and the meaning of
all other games is still open. And further more, if it's
acceptable for static game, is it possible to be developed into
Evolutionary Game Theory? And cooperative game? Is it related with
entangled system state?

As discussed in section \ref{phase}, because of the iteration
procedure and the distribution function we used, a natural way of
equilibrium calculation and refinement is provided by our
pseudo-dynamical method. The non-trivial phase transition
happening in Statistical Physics at finite $\beta = \beta_{c}$
implies the probability that when $N>>1$ the traditional
equilibrium state can be reached at some finite noise level, not
necessary at no-noise infinite-resolution background. In this
paper, we only argued such possibility, not by a real example.
Further analysis should be done to confirm such statement,
although we believe it from the background in Physics.

And as discussed in section \ref{qqgame}, when our representation
is used in Quantum Game Theory, a set of base vectors of strategy
(operator) space and their inner product need to be defined to
form them as a Hilbert space. Then all the other procedures are
quite straightforward. At least, it gives equivalent description.
But there are still some open questions, like what's the meaning
of a non-unitary operator in the strategy space? Does physical
operator have to be unitary operator? Another interesting question
is the effect of base vector transformation of Hilbert space. What
happens if base vectors other than our $\left( N^{c},F^{c}, N^{q},
F^{q}\right)$ are used?

We have to say our present result is a theory far from complete.
It stacks in our hands for a very long time, now we want to share
the idea with all. In fact, it's even possible to be nothing than
a toy representation of Game Theory. However, even in such case,
it's still of little value to provide a unified description and a
possible pseudo-dynamical equation theory which might be completed
later so that the end state of iteration from an arbitrary initial
will be the Equilibrium State. As you may already noticed our
paper is filled with questions other than their answers. Hopefully
it will motivate the discussion. Ironically, during the revision
of this paper, we found that the idea using a Hilbert space to
describe classical and quantum strategies has been proposed in
\cite{marinatto} long time before. So our works can be regarded as
a realization and development of this idea. In our paper, not only
strategies, but also payoff functions has been reexpressed into
Hilbert space and operators on it.

\section{Conclusions and outlook}
\label{conclusion}

Besides lots of questions in above section, here we summarize the
reliable conclusions we have till now. First, in the new
representation, all games including classical, quantum, even
entangled game, under general $N$-player
$\left(\prod_{i=1}^{N}L_{i}\right)$ case, can be defined by a
unified definition as equ(\ref{ourquantumgame}). All the
difference among the games is at the base vectors of strategy
space and the payoff matrix --- a $\left(1,1\right)$-tensor. In
the traditional form, payoff function of $N$-player classical game
is $\left(0,N\right)$-tensor; and for quantum game, it depends on
$\mathcal{H}$ and $\rho_{0}$, even not a tensor form. We have to
use special language for every specific game.

Second, in our representation, for quantum games, it's easy to see
the role of Off-diagonal elements of payoff matrix when we say a
quantum player can make more money over classical player. If the
payoff matrix is diagonal, it makes no difference, although a
quantum player can make use of quantum pure strategy, which has
off-diagonal elements in density matrix. Game Quantum is only
possible when both density matrix and payoff matrix have
off-diagonal elements. However, unfortunately, as we have pointed
out in section $\S\ref{spayoffmatrix}$, quantization of classical
game probably changes the definition of original classical game.

Third, with the form of payoff matrix and reduced payoff matrix,
equilibrium density matrix in Quantum Statistical Mechanics
$e^{\beta H_{R}^{i}}$ gives an equilibrium distribution over
strategy space. This provides some flexibility on the application
of game theory such as average behavior and collapse into Nash
Equilibrium under infinite resolution level ($\beta = \infty$).
This helps take partial rationality into our theory. Therefor, for
classical game, although our new representation brings nothing new
theoretically, it may provide some new technical tools to analyze
the NE, to include partial rationality, and to develop
evolutionary game.

If such a representation can provide some other insightful
advantage besides an equivalent representation of both Classical
and Quantum Game Theory, it's necessary to try more real games,
both classical and quantum, in the new framework. From the section
$\S$\ref{eigenvalue}, we see that because the system space is the
direct space of all players', the matrix form will be so large
that it make all calculations un-convenient. In Quantum Mechanics,
the idea to solve such problem is to introduce particle-number
representation to replace direct product of base vectors. For
undistinguishable particles, such approach significantly reduce
the hardwork of calculation. Maybe such simplification can be
generalized into Game Theory.

\section{Acknowledgement}
The author want to thank Dr. Shouyong Pei, Qiang Yuan, Zengru Di,
Yougui Wang  and Dahui Wang for their inspiring discussion and the
great patience to listen to my struggling argument at anytime.
Thanks also is given to Prof. Zhanru Yang, Dr. J. Eisert and Dr.
L. Marinatto for their suggestions to the revision.

\end{document}